\documentclass[times,tight]{aastex62}

\usepackage{amsfonts}
\usepackage{mathrsfs}
\usepackage{lineno}

\def\cblue{\color{blue}}

\def\sunm{M_{\odot}}
\def\SgrA{Sgr\,A$\!^{*}$\,}

\newcommand{\calJ}{{\cal J}}
\newcommand{\calD}{{\cal D}}
\newcommand{\calM}{{\cal M}}

\newcommand{\ergs}{{\rm erg\,s^{-1}}}
\newcommand{\RBon}{R_{\rm Bon}}
\newcommand{\Rg}{R_{\rm g}}

\newcommand{\mathdotM}{\dot{\mathscr{M}}}

\def\ihep{Key Laboratory for Particle Astrophysics, Institute of High Energy Physics,
Chinese Academy of Sciences, 19B Yuquan Road, Beijing 100049, China}
\def\AstroUCAS{School of Astronomy and Space Sciences, University of Chinese Academy of Sciences, 
19A Yuquan Road, Beijing 100049, China}

\def\PhyUCAS{School of Physics, University of Chinese Academy of Sciences, 
19A Yuquan Road, Beijing 100049, China}

\def\naoc{National Astronomical Observatory of China, 20A Datun Road, Beijing 100020, China}

\def\USTC{Department of Astronomy, University of Science and Technology of China, Hefei 230026, Anhui, China}

\def\kiaa{Kavli Institute of Astronomy and Astrophysics, Peking University, Beijing 100871, China}

\def\AstroPKU{Department of Astronomy, Peking University, Beijing 100871, China}

\received{\it 2023 August 24}
\accepted{\it 2023, November 11}

\shorttitle{Accretion-modified Stars in AGNs}
\shortauthors{Wang et al.}

\begin{document}

%\title{\large\bf %periodic variations of the near-infrared continuum of Sgr\,A$^{*}$: 
%Evidence for 
%an accreting stellar black hole orbiting around the central massive black hole in Sgr\,A$^{*}$}

\title{\large\bf
Accretion-modified Stars in Accretion Disks of Active Galactic Nuclei: the Low-luminosity Cases \vglue 0.15cm 
and an Application to \SgrA}% and Quasi-periodic Eruptions in Galactic Centers}

\author[0000-0001-9449-9268]{Jian-Min Wang}
\affil{\ihep}
\affil{\AstroUCAS}
\affil{\naoc}

\author{Jun-Rong Liu}
\affil{\ihep}
\affil{\PhyUCAS}

\author{Yan-Rong Li}
\affil{\ihep}

\author{Yu-Yang Songsheng}
\affil{\ihep}

\author{Ye-Fei Yuan}
\affil{\USTC}

\author{Luis C. Ho}
\affil{\kiaa}
\affil{\AstroPKU}

%\linenumbers

\begin{abstract}
In this paper, we investigate the astrophysical processes of stellar-mass black holes (sMBHs) 
embedded in advection-dominated accretion flows (ADAFs) of supermassive black holes (SMBHs) 
in low-luminosity active galactic nuclei (AGNs). The sMBH is undergoing Bondi accretion at a 
rate lower than the SMBH. Outflows from the sMBH-ADAF dynamically interact with their surroundings 
and form a cavity inside the SMBH-ADAF, thereby quenching the accretion onto the sMBH. Rejuvenation 
of the Bondi accretion is rapidly done by turbulence. These processes give rise to quasi-periodic 
episodes of sMBH activities and creat flickerings from relativistic jets developed by the 
Blandford-Znajek mechanism if the sMBH is maximally rotating. Accumulating successive sMBH-outflows 
trigger viscous instability of the SMBH-ADAF, leading to a flare following a series of flickerings. 
Recently, the similarity of near-infrared flare's orbits has been found by GRAVITY/VLTI astrometric 
observations of Sgr A$\!^{*}$: their loci during the last 4-years consist of a ring in agreement 
with the well-determined SMBH mass. We apply the present model to \SgrA, which shows quasi-periodic 
flickerings. An sMBH of $\sim 40\sunm$ is preferred orbiting around the central SMBH of \SgrA from 
fitting radio to X-ray continuum. Such an extreme mass ratio inspiraling (EMRI) provides an excellent 
laboratory for {\it LISA\,/\,Taiji\,/\,Tianqin} detection of mHz gravitational waves with strains of 
$\sim 10^{-17}$, as well as their polarization.
\end{abstract}
\subjectheadings{Active galactic nuclei (16); Galaxy accretion disks (562); Supermassive black holes (1663)}

\section{Introduction}
The model of accretion onto a single supermassive 
black hole (SMBH) is successful to explain the powerful radiation of active galactic nuclei (AGNs) 
\citep[e.g.,][]{Rees1984}, however, there is a growing body of evidence suggesting that some new 
ingredients should be incorporated into this canonical model.
Star formation has been suggested by many authors for different purposes since the early attempts 
of \cite{Kolykhalov1980} in light of the self-gravity of outer parts of the AGN accretion 
disks \citep{Paczynski1978}. Consequently,
it may be an efficient way of fueling gas to galactic centers, triggering the activity of
SMBHs \citep{Shlosman1989,Thompson2005,Wang2010}. Spectral energy distributions 
of AGNs could be revised by star formation \citep{Goodman2003,Goodman2004,Thompson2005} as well as the
origins of high metallicity \citep{Collin1999,Collin2008,Wang2011,Wang2012,Grishin2021,Wang2023,Fan2023} 
observed 
in AGN broad-line regions \citep{Hamann1998,Warner2003,Nagao2006,Shin2013,Du2014}. After supernovae
explosions of stars formed in the AGN disks, there remain compact objects as satellites of the central 
SMBHs. Obvious questions remain 
as to the fates of the satellites embedded in the disks, what is their observational appearance?

The potential association of quasar SDSS J1249+3449 with GW190521 \citep{Graham2020}, consisting of 
$(85+66)\sunm$ binary BHs, by the Advanced LIGO/Virgo consortium \citep{Abbott2020} was motivated by
the formation of such a massive BH binary in special environments. See more candidates of potential 
associations of LIGO gravitational wave detection with quasars \citep{Graham2023}. These mergers of 
high-mass stellar black hole binaries are much beyond productions from the evolution of 
stars \citep{Woosley2002}, indicating that they are formed in a very dense environment. They have 
stimulated renewed interest in the question of the fates of compact objects in the AGN disks 
\citep[e.g.,][]{McKernan2012,Bellovary2016,Bartos2017,Stone2017,Secunda2019,Yang2019,Tagawa2020,Graham2020,
Wang2021a,Wang2021b,Samsing2022}. 
This interesting idea can be traced back to the early paper of \cite{Cheng1999}, who suggest that compact 
objects formed in AGN disks undergo mergers generating $\gamma$-ray bursts and gravitational 
waves. The terminology of accretion-modified stars (AMS) used in this series are referred to as ones 
with accretion from AGN disks \citep{Wang2021a}, where the accreting objects could be main sequence 
stars \citep{Wang2023}, stellar-mass black holes (sMBHs) \citep{Wang2021b}, neutron stars \citep{Zhu2021a}, 
or white dwarf stars \citep{Zhu2021b,Zhang2023}. AMS phenomena exhibit distinguished features in different environments as predicted by the above papers.

This is the third paper of the series exploring the observational signatures of accreting black holes
in contexts of AGN accretion disks \citep{Wang2021a,Wang2021b}. The sMBHs have been studied in the 
environment of standard accretion disks \citep{Wang2021a} showing electromagnetic signatures of the
Bondi explosion, and Jacobi capture (sMBHs are captured by each other through their tidal forces in 
nearby orbits around the central SMBHs) \citep{Wang2021b}. This paper focuses on studying the 
signatures of sMBH embedded in advection-dominated accretion flows (ADAFs) of low luminosity AGNs 
(LLAGNs). Quasi-periodic flickerings are suggested to occur over the whole electromagnetic wave bands. 
Flares following a series of flickerings are produced by accumulated energies of the sMBH-outflows. 
We apply the present model to  Sagittarius A$\!^{*}$
(\SgrA) for its variabilities and multiwavelength emissions in \S\ref{sec:application}. 
\SgrA is an excellent laboratory of milli-Hertz (mHz) gravitational waves.

\begin{figure*}
    \centering
\includegraphics[height= 0.285\textwidth,trim=45 100 50 85,clip]{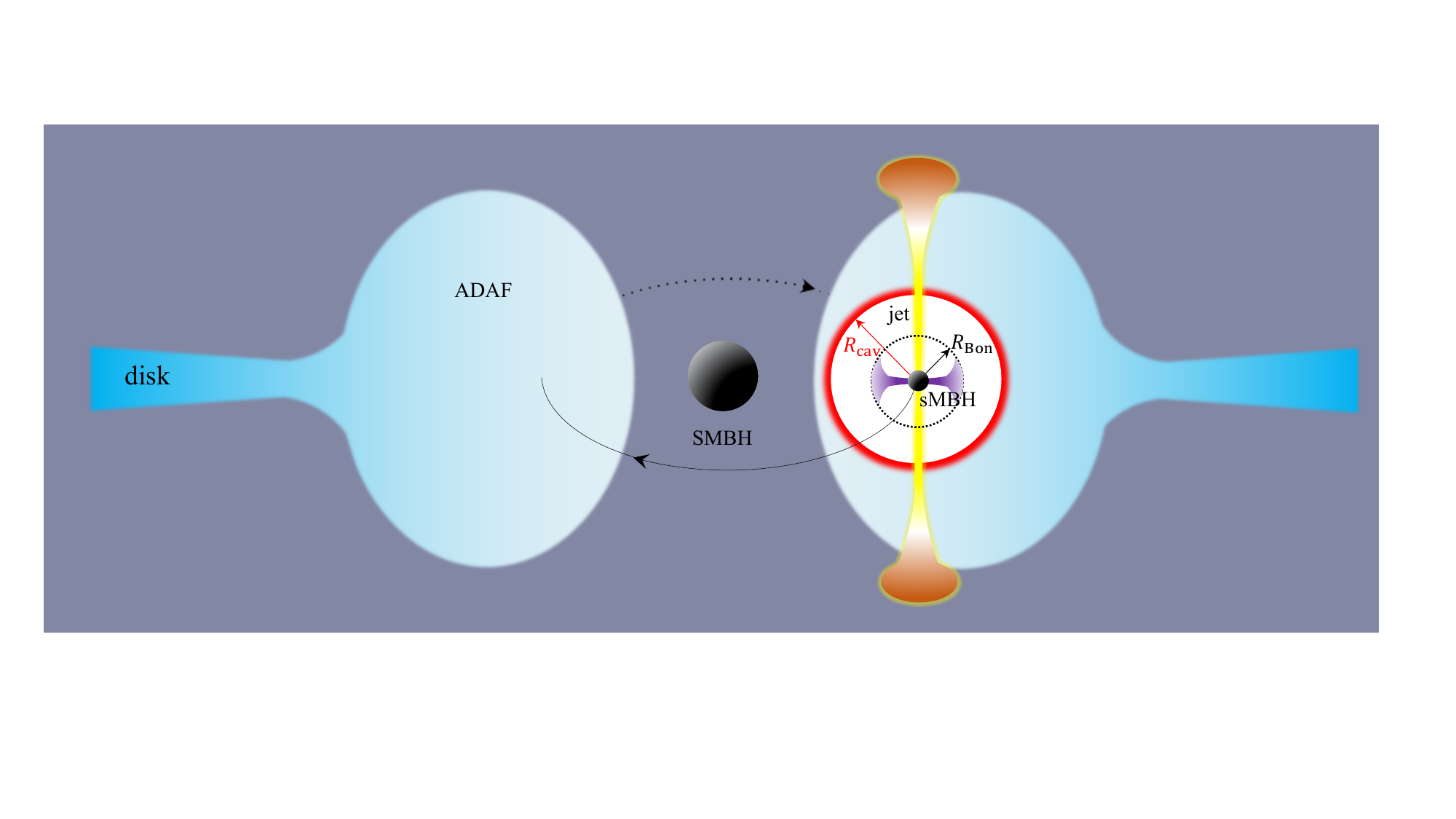}
\includegraphics[height= 0.285\textwidth,trim=20 15 25 33,clip]{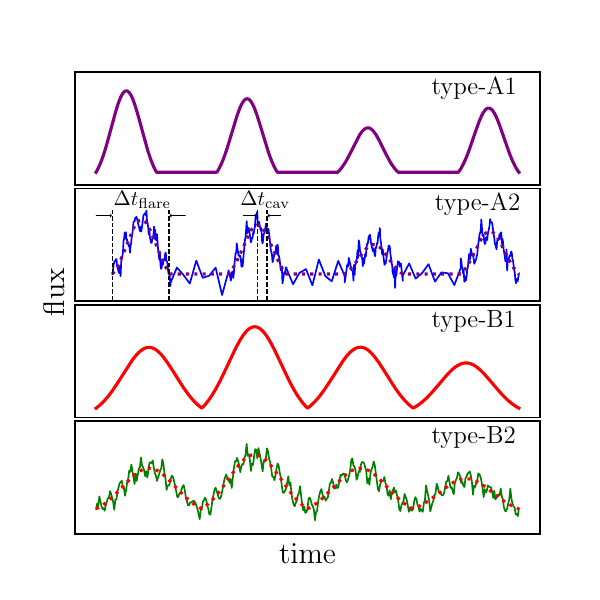}
    \caption{\footnotesize {\it Left}:
    An sMBH embedded in SMBH-ADAFs is orbiting around the central SMBH. 
    The sMBH is accreting with Bondi rates from the SMBH-ADAF, but outflows 
    from the sMBH-ADAFs drive the formation of a cavity and quench the accretion. Such 
    an accretion-feedback system exhibits quasi-periodic behaviors, showing quasi-periodic flickerings
    from relativistic jets produced through the Blandford-Znajek (BZ) mechanism of the sMBH-ADAF.
    Emissions of the jet constitute spectral energy distributions additional to the SMBH-ADAF.
    The jet may be partially choked by the SMBH-ADAF, giving rise to sub-relativistic wisps from the
    SMBH-ADAF. The accumulating sMBH-outflows will trigger the viscosity instability of the SMBH-ADAF 
    resulting in
    flares following a series of flickerings. The present model can explain the NIR quasi-periodic 
    flickerings and flares of \SgrA. {\it Right:} Illustrated classifications of 
    light curves. Type-A shows that flares/flickerings are 
    superposed on a stationary state whereas type-B shows an upon-down mode. Flickerings are
    thin lines and flares are thick lines.
        }
\label{fig:model}
\end{figure*}

\section{The model}
Figure \ref{fig:model} shows the model of the present AMS.
An sMBH (the secondary BH as one satellite of the SMBH) is orbiting around the central SMBH (the primary) 
inside the ADAF. 
We assume a circular orbit in this binary black hole system. Parameters with the subscripts of 
``s'' and ``p'' are referred to as those of the secondary and the primary.
Their dimensionless accretion rates are defined by 
$\mathdotM_{\rm p,s}=\dot{M}_{\rm p,s}/\dot{M}_{\rm Edd}^{\rm p,s}$, where $\dot{M}_{\rm p,s}$ are the 
accretion rates, $\dot{M}_{\rm Edd}^{\rm p,s}=L_{\rm Edd}^{\rm p,s}c^{-2}$ is their corresponding limit 
rate in light of the Eddington luminosity  
$L_{\rm Edd}^{\rm p,s}=4\pi GM_{\rm p,s} m_{\rm p}c/\sigma_{\rm T}$, where $G$ is the gravitational constant,
$c$ is the speed of light, $m_{\rm p}$ is the proton mass, and $\sigma_{\rm T}$
is the Thomson cross-section. When $\mathdotM$ is much less than the unity, the accretion flows
are supported by the ion pressure of the plasma with two temperatures because cooling is so 
inefficient that most of the released energies are not able to radiate away. This results in that 
the accretion flows are very hot and are able to produce relativistic
jet \citep{Rees1982}. This is the early version of ADAF. For simplicity, we use the 
self-similar solution of ADAF for the primary BH \citep{Narayan1995}, in which 
the half-thickness, density, and sound velocity are 
\begin{equation}\label{Eq:disk}
\left\{
\begin{array}{lll}%\vspace{0.5ex}
H_{\rm p}&=&1.2\times 10^{12}\,M_{6}r_{1}\,{\rm cm},\\ %\vspace{0.5ex}
n_{\rm e}&=&9.3\times 10^{9}\,\alpha_{0.1}^{-1}M_{6}^{-1}\mathdotM_{{\rm p},\bar{3}}r_{1}^{-3/2}\,{\rm cm^{-3}},\\ %\vspace{0.5ex}
c_{\rm s,p}&=&5.0\times 10^{9}\,r_{1}^{-1/2}\,{\rm cm\,s^{-1}},\\ %\vspace{0.5ex}
v_{\rm r}&=&3.9\times 10^{8}\,\alpha_{0.1}r_{1}^{-1/2}\,{\rm cm\,s^{-1}},
\end{array}\right.
\end{equation}
where $\alpha_{0.1}=\alpha/0.1$ is the viscosity parameter \citep{Shakura1973}, $r_{1}=R_{\bullet}/10\Rg$ 
is the radius of the disk from the SMBH, and $\Rg=GM_{\rm p}/c^{2}$ is the gravitational radius, 
$\mathdotM_{\rm p,\bar{3}}=\mathdotM_{\rm p}/10^{-3}$, $M_{6}=M_{\rm p}/10^{6}\sunm$
is the mass of the primary black hole. There are three constants of $c_{1,2,3}$ in the self-similar
solution \citep{Narayan1995}. In the case of an advection fraction ($f_{\rm adv}=0.9$ and the 
adiabatic index of
$\gamma=4/3$), we calculate the three constants of $c_{1,2,3}=(0.42,0.55,0.28)$ from their
definitions in this paper. Though the self-similar solution is used here, it is in 
agreement with the numerical solutions of the global ADAF.
The density given by Eq.\,(\ref{Eq:disk}) is consistent with {\it Chandra} X-ray observations
of \SgrA, where the column density $N_{\rm H}\approx 8.0\times 10^{22}\,{\rm cm^{-2}}$ at the
center \citep{Wang2013}, which is consistent with the inward extrapolation of the gas density from the
Bondi radius \citep{Xu2006}. The ADAF scattering optical depth is about
$\tau_{\rm es}\approx 0.007$ at $10\,\Rg$ in the vertical
direction for these typical values given by Eq.\,(\ref{Eq:disk}), leading to different situations from
the Shakura-Sunyaev disks, the jet developed by the sMBH could be not completely choked. 
The most prominent feature of the ADAF is the positive Bernoulli constant (defined as
a sum of kinematic, thermal and potential energies of the ADAF gas),
implying that outflows are developed by the advection mechanism since the accretion flows 
become gravitationally unbound by the SMBH \citep{Narayan1994}. 
%{\cblue Event Horizon Telescope (EHT) observations of \SgrA show the density and temperatures
%of the ADAF \citep{AAA} in \SgrA are consistent with these typical values given by Eq.(\ref{Eq:disk}).}
Though the radiative efficiency of the ADAF is very low, the efficiency of outflows is quite high, resulting 
in strong feedback to the SMBH-ADAF and making a cavity around the sMBH.

For simplicity, we assume that the sMBH is trapped inside the SMBH-ADAF, and its Bondi accretion rates are 
given by
\begin{equation}\label{eq:Bondirate}
\mathdotM_{\rm s}=\frac{\dot{M}_{\rm s}}{\dot{M}_{\rm Edd}}
                 =\frac{4\pi G^{2}M_{\rm s}^{2}\rho_{\rm p}}{c_{s}^{3}\dot{M}_{\rm Edd}}
                 =2.0\times 10^{-6}\,\alpha_{0.1}^{-1}\mathdotM_{\rm p, \bar{3}}q_{\bar{5}},
%                 \left(\frac{M_{6}}{4}\right)^{-1}\left(\frac{m_{1}}{4}\right),
\end{equation}
where $M_{\rm s}$ is the mass of sMBH, $q_{\bar{5}}=q/10^{-5}$, $q=M_{\rm s}/M_{\rm p}$ is
the mass ratio of the sMBH to the central SMBH, and $\rho_{\rm p}=n_{e}m_{\rm p}$ is the mass density
of the SMBH-ADAF. This indicates that the sMBH is also
the status of ADAF. It is interesting to note that the sMBH accretion rates are independent of
its location. The Bondi radius of the sMBH
\begin{equation}
\RBon=\frac{GM_{\rm s}}{c_{\rm s,p}^{2}}=2.1\times 10^{8}\,\left(\frac{m_{1}}{4}\right)r_{1}\,{\rm cm},
\end{equation}
where $m_{1}=M_{\rm s}/10\,\sunm$ is the sMBH mass. $R_{\rm Bon}$ is much smaller than the Hill radius of
$R_{\rm Hill}=(M_{\rm s}/M_{\rm p})^{1/3}R_{\bullet}
  \approx 3.2\times 10^{10}\,m_{1}^{1/3}M_{6}^{2/3}r_{1}\,{\rm cm}$,
the SMBH tidal disruption of 
the AMS can be avoided. On the other hand, the outer radius (or the circularized radius) 
of the sMBH-ADAF can be estimated
by angular momentum balance. The net specific angular momentum (AM) of the SMBH-ADAF is given by
$\Delta \ell_{\rm p}=\Delta R\sqrt{GM_{\rm p}/R_{\bullet}}$, where $\Delta R$ is the width of the belt at $R$.
The specific AM of the sMBH-ADAF can be approximated by 
$\Delta \ell_{\rm s}=\sqrt{X_{\rm out}GM_{\rm s}}$, where $X_{\rm out}$ is the outer radius of the 
sMBH-ADAF. Equating $\Delta \ell_{\rm p}=\Delta \ell_{\rm s}$ and $\Delta R\approx X_{\rm out}$, we have
approximately 
$X_{\rm out}=\left(M_{\rm s}/M_{\rm p}\right)R_{\bullet}=1.5\times 10^{7}\,q_{\bar{5}}r_{1}M_{6}\,$cm,
which is consistent with numerical simulations \citep[e.g.,][]{Igumenshchev1999}. 
The $X_{\rm out}$ is significantly smaller than the Bondi radius. 

It should be noted that the orbiting sMBH suggested here is rotating with a supersonic velocity
($v_{\rm rot}\sim c/3$) at $10\,\Rg$. Shocks due to the motion could be produced to accelerate 
some electrons. Subsequently, non-thermal emissions of the electrons are thus radiated from this 
region. A similar case of the Bondi sphere of an isolated black hole supersonically moving in 
medium has been discussed by \cite{Wang2014}, however, we will not discuss this potentially important
point here. In this paper, only the case with an extremely low-$q$ system is considered. Such a 
system allows us to use the perturbation approximation to treat the influence of the sMBH on the
SMBH-ADAF. Otherwise, we have to consider the inhomogeneity of the SMBH-ADAF caused by the
sMBH feedback. We consider that the SMBH-ADAF is in a stationary state.

\subsection{Cavity and flares}
Since the progenitor of the sMBH should rotate very fast, the sMBH should also rotate fast
and undergo two processes to generate influence on the SMBH-ADAF. First, energetic outflows with a
power of $L_{\rm out}=\eta_{\rm out}\dot{M}_{\rm s}c^{2}$ are produced since the Bernoulli constant 
is positive, where $\eta_{\rm out}$ is the conversion efficiency of channeling gravitational energy 
into outflows. We note that $\eta_{\rm out}$ is much higher than the radiative efficiency of the ADAF
and take $\eta_{\rm out}=0.1$ as a typical value in this paper. The higher $\eta_{\rm out}$ the more 
prominent effects of the sMBH on the SMBH-ADAF. Second, the Blandford-Znajek processes 
\citep[BZ:][]{Blandford1977} form bi-polar relativistic jets. The outflows from sMBH-ADAF heat 
gas within the Bondi radius are efficiently clearing the gas in a timescale of 
$\Delta t_{\rm Bon}=\Delta E_{\rm Bon}/L_{\rm out}$, where 
$\Delta E_{\rm Bon}=GM_{\rm s}\Delta M_{\rm Bon}/R_{\rm Bon}$ is the gravitational energy 
of the gas, $\Delta M_{\rm Bon}\approx \frac{4\pi}{3}R_{\rm Bon}^{3}n_{e}m_{\rm p}$
is the gas mass within $R_{\rm Bon}$. For $n_{e}=10^{10}\,{\rm cm^{-3}}$, we have 
$t_{\rm Bon}\sim 0.017\,$s,
and this demonstrates the sMBH-outflows have efficient feedback to form a cavity in the SMBH-ADAF.
On the other hand, the smearing timescale of the Bondi cavity is approximately  
$R_{\rm Bon}/\alpha c_{\rm s}\sim 10^{-3}\,$s. In the Appendix, we list the other two classes 
of cavity formation. They have timescales much longer
than $t_{\rm Bon}$, but significantly shorter than the flickerings and flares in \SgrA. 
Moreover, the cooling timescale of local SMBH-ADAF is much longer than these cases.
It can be regarded as a successive series of these events that creat a cavity as described below.

Outflows from the sMBH-ADAF will significantly affect the local structures of the SMBH-ADAF, 
generating a cavity with a radius of $R_{\rm cav}$, provided that the outflow kinetic energies 
exceed the local dissipation rates of gravitational energy of the SMBH-ADAF. This condition can 
be expressed as
\begin{equation}\label{eq:Lout}
L_{\rm out}\,{\cblue \gtrsim}\, \frac{4\pi}{3} R_{\rm cav}^{3}\varepsilon_{+},
\end{equation}
where $\varepsilon_{+}\approx Q_{+}/H_{\rm p}$ is the volume dissipation rates of the SMBH-ADAF, 
$Q_{+}=3GM_{\rm p}\dot{M}_{\rm p}/8\pi R_{\bullet}^{3}$ is surface rates \citep{Frank2002}. 
Here, we neglect the factor of the inner boundary condition of accretion disks. We would 
like to point out the implications of Eq.\,(\ref{eq:Lout}), i.e., the outflow energies are the 
extra source of the SMBH-ADAF as type A variability shown in Fig.\,\ref{fig:model}. 
In the following discussions, we take the equal form of inequality (\ref{eq:Lout}), 
and some results are the upper and lower limits for a given $L_{\rm out}$, for instance, 
$R_{\rm cav}$ in Eq.\,(\ref{eq:Rcav}) and $\Delta t_{\rm cav}$ in (\ref{eq:Deltatcav}), 
respectively. The cavity can be made by the work done within a time interval $\Delta t_{\rm cav}$ 
through the outflow-driven expansion of the heated part
\begin{equation}\label{eq:Loutcav}
L_{\rm out} \Delta t_{\rm cav}=\frac{4\pi}{3}R_{\rm cav}^{3}P_{\rm gas},
\end{equation}
where $P_{\rm gas}=\rho_{\rm p}c_{\rm s}^{2}$ is the gas pressure of the SMBH-ADAF.
Actually,  Eq.\,(\ref{eq:Loutcav}) shares the same meaning of the work done 
$P\Delta V$ of AGN feedback in galaxy cluster \citep[e.g.,][]{McNamara2007,Fabian2012}.
The ram pressure of the sMBH-outflows impedes the
inflows outside the cavity through the balance with the surrounding medium, namely
$L_{\rm out}/v_{\rm out}=4\pi R_{\rm cav}^{2}P_{\rm gas}$.
We find that this condition holds as long as the outflows have a Mach number of 
$\calM=v_{\rm out}/c_{\rm s}\ge 1$.  Eq.\,(\ref{eq:Loutcav}) describes the working 
process of cavity formation, and is independent of (\ref{eq:Lout}).
In the Appendix, we discuss the other two possibilities 
of cavity formation and find that they naturally satisfy the conditions (Eq.\,\ref{eq:Lout} 
and \ref{eq:Loutcav}) employed here. During the cavity formation, we approximate the gas 
pressure as a constant. From this energy budget, we can derive the cavity radius
\begin{equation}\label{eq:Rcav}
R_{\rm cav}=\left(\frac{3L_{\rm out}H}{4\pi Q_{+}}\right)^{1/3}
           =1.9\times 10^{10}\,\eta_{\rm 0.1}^{1/3}\alpha_{0.1}^{-1/3}r_{1}^{4/3}
            q_{\bar{5}}^{2/3}\left(\frac{M_{6}}{4}\right)\,{\rm cm},
\end{equation}
and the time for cavity formation
\begin{equation}\label{eq:Deltatcav}
\Delta t_{\rm cav}=\frac{HP_{\rm gas}}{Q_{+}}
        =45.9\,\alpha_{0.1}^{-1}r_{1}^{3/2}\left(\frac{M_{6}}{4}\right)\,{\rm min},
\end{equation}
where $\eta_{0.1}=\eta_{\rm out}/0.1$. $R_{\rm cav}$ is comparable to the Hill 
radius, but is still much smaller than the half-thickness of the SMBH-ADAF. In the case of
the SMBH-ADAF, $R_{\rm cav}$ is independent of the accretion rates of the SMBH-ADAF, but 
dependent of the location and mass of the sMBH, and the SMBH mass. Since 
$R_{\rm cav}\gg R_{\rm Bon}$, the sMBH will stop accretion in the timescale of the cavity
formation. It is very interesting to note that the formation timescale is independent of
the sMBH and its outflows but sensitive to its location radius and the SMBH mass. This 
offers an opportunity to measure the viscosity as the hardest parameter of accretion 
macro-physics if the location, sMBH mass, and flickering periods are fixed in the future 
(after detections of mHz gravitational waves). Actually, this $\Delta t_{\rm cav}$ is just 
the thermal instability timescale of 
$t_{\rm therm}=\Sigma c_{\rm s}^{2}/Q_{+}$ \citep[e.g.,][]{Frank2002}. 
We get this from the energy budget avoiding details of the expansion dynamics. Simultaneously, 
we should note that the BZ process generates relativistic ejecta during $\Delta t_{\rm cav}$. 
This produces flickerings as a result of the quenched Bondi accretion of the sMBH-outflows. 

On the other hand, this cavity could be slacked down by the dynamical interaction or turbulence 
after the cessation of outflows. This rejuvenates the Bondi accretion of the sMBH. For a simple 
estimation, we can estimate the timescale of a rejuvenation
\begin{equation}
\Delta t_{\rm rej}=\frac{R_{\rm cav}}{v_{\rm tur}}=37.9\,\eta_{0.1}^{1/3}\alpha_{0.1}^{-4/3}
            q_{\bar{5}}^{2/3}r_{1}^{11/6}\left(\frac{M_{6}}{4}\right)\,{\rm s},
\end{equation}
where $v_{\rm tur}=\alpha c_{\rm s}$ is the turbulence velocity,
which is much shorter than $\Delta t_{\rm cav}$. The cavity is then destroyed 
by the turbulence of the cavity developed by the interaction between the outflows and the SMBH-ADAF.
It is therefore expected that the cavity appears periodically with a timescale of 
$\Delta t_{\rm cav}+\Delta t_{\rm rej}\approx \Delta t_{\rm cav}$. A flickering lasts 
for $\Delta t_{\rm cav}$. It is very interesting to note that 
$\Delta t_{\rm cav}$ only sensitively depends on the location of the sMBH given the binary masses.
From the observational side, the flickering period can be used to estimate the location.
In practice, if the density of the SMBH-ADAF could be inhomogeneous, the flickering timescale could
vary at different epochs. On the averaged behaviors, the sMBH undergoes quasi-periodic
activities appearing as quasi-periodic flickerings. This is a unique feature of the AMS inside 
SMBH-ADAFs.

The sMBH is undergoing episodic Bondi accretion, but the cavity continually grows with time since 
the SMBH-ADAF is cooling much slower than the outflow-driven heating of the sMBH-ADAF. Therefore, 
the cavity density decreases, but its temperature and radius increase with time. 
Cooling processes inside the SMBH-ADAF involves bremsstrahlung (proton-electron and electron-electron 
collisions), synchrotron radiation and inverse Compton scattering (IC) \citep[e.g.,][]{Narayan1995}. 
Magnetic fields are usually estimated by the magnetization factor $\beta$, which is defined by
$P_{\rm gas}=\beta(P_{\rm gas}+P_{\rm mag})$, where $P_{\rm mag}=B^{2}/8\pi$ is the magnetic pressure
and $B$ is magnetic field. For the 
case of $\beta=0.5$, $n_{e}=10^{10}\,{\rm cm^{-3}}$ and $T_{\rm e}=10^{9}\,$K, we obtain the total 
emissivity of $\epsilon_{\rm tot}\approx \chi\,\epsilon_{\rm ff}$ and $\chi\approx 28$ (this can also 
be justified by comparing the NIR and hard X-ray peaks of ADAF SED as shown in Figure\,\ref{fig:sed_mod}), 
where $\epsilon_{\rm ff}$ is the free-free emissivity. The cooling factor $\chi$ depends on the 
density, temperature and magnetization factor $\beta$. The cooling 
timescale of the shocked gas 
\begin{equation}\label{eq:tcool}
\Delta t_{\rm cool}=\frac{3}{2}\frac{n_{\rm e}kT_{\rm s}}{\epsilon_{\rm tot}}
            \approx 5.6\,\chi_{0}^{-1}T_{9}^{1/2}n_{10}^{-1}\,{\rm hour},
\end{equation}
where $\chi_{0}=\chi/28$, $n_{10}=n_{e}/10^{10}\,{\rm cm^{-3}}$, $k$ is the Boltzmann constant 
and $T_{9}=T_{\rm s}/10^{9}\,$K is the 
temperature of the shocked SMBH-ADAF. Since $\Delta t_{\rm cool}\gg \Delta t_{\rm cav}$,
the SMBH-ADAF gas will be continuously heated by the sMBH-ADAF outflows, 
and the cavity grows with time. When it reaches the cooling timescale,
the sMBH outflow accumulates power enough to produce a flare with a cavity radius 
\begin{equation}\label{eq:Rflare}
R_{\rm flare}\approx
     \left(\frac{3 L_{\rm out}\Delta t_{\rm cool}}{4\pi P_{\rm gas} }\right)^{1/3}
             =3.7\times 10^{10}\,\eta_{0.1}^{1/3}\chi_{0}^{-1/3}T_{9}^{1/6}n_{10}^{-1/3}
             r_{1}^{5/6}q_{\bar{5}}^{2/3}\left(\frac{M_{6}}{4}\right)^{2/3}\,{\rm cm},
\end{equation}
from Eq.(\ref{eq:Loutcav}), which is still much smaller than the thickness of the SMBH ADAF.
The viscous instability of the SMBH-ADAF is developing in a timescale of 
$\Delta t_{\rm vis}\approx \alpha^{-1}(H/R_{\bullet})^{-2}t_{\rm \phi}$, where $t_{\phi}$
is the orbit period \citep[e.g.,][]{Frank2002}. For typical values and $H/R_{\bullet}\sim 1$, we have 
$\Delta t_{\rm vis}\approx 11.0\,\alpha_{0.1}^{-1}(M_{6}/4)r_{1}^{3/2}\,$hour, which is comparable 
with the cooling timescale. Therefore, the viscous instability will be unavoidably triggered 
by the accumulated energies of the sMBH-ADAF outflows. A flare releases 
the total energies of
\begin{equation}
E_{\rm flare}\approx \Delta t_{\rm flare}Q_{+}\left(\pi R_{\rm flare}^{2}\right)
             \approx 1.5\times 10^{40}\,\eta_{0.1}^{2/3}\chi_{0}^{-5/3}
             T_{9}^{5/6}n_{10}^{-5/3}r_{1}^{-4/3}\mathdotM_{\rm p,\bar{3}}
             q_{\bar{5}}^{4/3}\left(\frac{M_{6}}{4}\right)^{1/3}\,{\rm erg},
\end{equation}
where $\Delta t_{\rm flare}=\min(\Delta t_{\rm vis},\Delta t_{\rm cool})$ and we take
$\Delta t_{\rm flare}=\Delta t_{\rm cool}$ here for a simple treament. A flare is rising 
with $\Delta t_{\rm flare}$, and decaying with $\Delta t_{\rm cool}$. The present model predicts 
that flares happen at a timescale of a few hours ($\Delta t_{\rm flare}$), and a few flares per 
day. %but $\Delta t_{\rm flare}$ can be used as the characteristic variability timescales. 
However, the quiescent phase ($\Delta t_{\rm quiet}$) is complicated by the recovery of 
the flaring cavity of the SMBH-ADAF, which is controlled mainly by local cooling, viscosity-driven 
infalling gas of the local flows and viscosity dissipations of the gravitational energies.
Flares are not periodic because of uncertainties of $\Delta t_{\rm quiet}$. 
This unique feature can be
used to test the flickering origins (e.g., magnetic reconnection model). As we can see,
this is in agreement with the observed flares in \SgrA. For a flare state, the recovery of the
SMBH-ADAF returning to its thermal equilibrium depends on joint processes of cooling and dynamical
mixing, rather than a single process. Perturbation approximation is not valid in the state.

We would like to emphasize that the current treatments consider the sMBH activity 
as a perturbation. The validity of this approximation can be guaranteed provided that 
the expansion velocity of the cavity formation is sub-relativistic.
When the mass ratio is large enough, the formation of a cavity will be 
different from the present.  In such a case, the tidal torque of the sMBH-SMBH 
binary system is strong enough to engulf the SMBH-ADAF. This is very similar to the case of
exoplanet formation. The accretion rates of the sMBH are much 
lower than Eq.\,(\ref{eq:Bondirate}) since the SMBH-ADAF density should be replaced by
the density inside the gulf, however, it still significantly radiates for observations
(since sMBH is large).

\subsection{Quasi-periodic flickerings}
%Relativistic ejecta and jet}
%
Except for outflows from the sMBH-ADAF, relativistic ejecta would be developed by the 
BZ mechanism through extracting spin energy of the sMBH if it is rotating fast enough \citep{Blandford1977}.  
Given a BH with spin AM $\calJ_{\bullet}$, the pumping power is given \citep{Ghosh1997,Macdonald1982} 
\begin{equation}\label{eq:LLBZ}
L_{\rm BZ}=\left(\frac{1}{32}\right)\omega_{\rm F}^{2}B_{\perp}^{2}R_{\rm h}^{2}c
           \left(\frac{\calJ_{\bullet}}{\calJ_{\rm max}}\right)^{2},
\end{equation}
where $B_{\perp}$ is magnetic field $B$ normal to the horizon at $R_{\rm h}$,
$\calJ_{\rm max}$ is the maximum of the spin AM, 
$\omega_{\rm F}=\Omega_{\rm F}(\Omega_{\rm h}-\Omega_{\rm F})/\Omega_{\rm h}^{2}$ is the factor 
describing relative angular velocity of magnetic field to the BH ($\Omega_{\rm h}$). Large scale 
magnetic field of the sMBH-ADAF, $B_{\perp}=B_{\rm disk}(H/r_{\bullet})^{1/2}$, $r_{\bullet}$ is
the radius of the sMBH disk, which is formed by the fast 
radial motion \citep{Livio2003,Cao2011}, will be involved to form a jet, where 
$B_{\rm disk}$ is the azimuthal magnetic fields generated by the dynamo viscosity. Generally, 
geometrically thin disks $(H/r_{\bullet}\sim 10^{-2}-10^{-3})$ are not able to produce relativistic 
jets because of $B_{\perp}\ll B_{\rm disk}$. It should be noted that the sMBH-ADAF has much
stronger $B_{\rm disk}$ than that of the SMBH-ADAF, which can be justified by Eq.(2.15) in
\cite{Narayan1995}. For an optically thin ADAF, \cite{Armitage1999} 
calculated the BZ power 
\begin{equation}\label{eq:LBZ}
L_{\rm BZ}\approx\frac{\sqrt{14}}{192}j_{\bullet}^{2}\dot{M}_{\rm s}c^{2}
          =1.92\times 10^{32}\,\alpha_{0.1}^{-1}j_{\bullet}^{2}\mathdotM_{\rm p,\bar{3}}
           q_{\bar{5}}^{2}
          %\left(\frac{m_{1}}{4}\right)^{2}\left(\frac{M_6}{4}\right)^{-1}
          \left(\frac{M_6}{4}\right)\,{\ergs},
\end{equation}
where $j_{\bullet}=\calJ_{\bullet}/\calJ_{\rm max}$, and we use the Bondi accretion rates 
(Eq.\,\ref{eq:Bondirate}). Here, we take accretion rates of the sMBH during the interval
of $\Delta t_{\rm cav}$ as a constant as an averaged one given by Eq.\,(\ref{eq:Bondirate}). 
Temporal profiles of the $L_{\rm BZ}$ depend on the evolution of density and temperature 
of the cavity, we will investigate this issue in the future.

The observational appearance of the BZ power release depends on two classes of factors: 1) the 
proceeding of the cavity formation driven by episodic accretion onto sMBH
and 2) the dynamical interaction between the jet and the 
SMBH-ADAF, both of which determine the temporal profiles of the flickerings. 
These temporal processes make it very difficult to estimate the time-dependent Lorentz factor
of the jets. Considering the difficulties
%
%
%Additionally, the vertical structures of the ADAF are not homogeneous. In this paper, we 
%simplified the estimation of the kinematics. Eq. (14) gives the averaged Lorentz factor, where the 
%``averaged'' includes temporal and spatial aspects.
%
%Moreover, we find that this jet is slowed down at $H_{\rm p}/3$ from relativistic states to 
%sub-relativistic ones. This consideration relies on a fact that the jets are invisible flickering
%if their Lorentz factors are too small. The option of $H_{\rm p}/3$ is reasonable for this.
%
%The jet can penetrate the ADAF but it may emerge from the ADAF as a sub-relativistic ejecta. 
%Some ejecta (or wisps) have been observed by Rauch et al.(2016) and Middelberg et al.(2004), which
%are cited in our manuscript.}
%
in this paper, we consider 
two possible outcomes for the relativistic ejecta: 1) they are partially choked and exhibit 
sub-relativistic wisps after emerging; 2) they can penetrate the entire SMBH-ADAF resulting 
in the appearance of superluminal blobs. Without details of the jet dynamics, 
a flickering apparently appears 
when a jet can penetrate the SMBH-ADAF through a length of the ADAF and the Lorentz factor reaches 
\begin{equation}\label{eq:Gamma}
\Gamma\approx\frac{L_{\rm BZ}\Delta t_{\rm cav}}{\Delta M_{\rm j}c^{2}}
      \approx 8.5\,\alpha_{0.1}^{-1}x_{\rm j,1}^{-2}j_{\bullet}^{2}
       r_{1}^{2}\left(\frac{\ell_{j}}{H_{\rm p}/3}\right)^{-1},
\end{equation} 
where $\Delta M_{\rm j}=\pi X_{\rm b}^{2}\ell_{\rm j}n_{\rm e}m_{\rm p}$ is the mass of 
the jet, $x_{\rm j,1}=X_{\rm b}/10r_{\rm g}$ is the jet radius obtained from the analytical 
solution of the BZ process \citep[e.g.,][]{Chen2021}, $r_{\rm g}$ is the gravitational 
radius of the sMBH, and $\ell_{\rm j}$ is its length. This is the average Lorentz factor 
of the jet, which is similar to that of radio-loud AGNs \citep{Ghisellini1993}. 
After the Lorentz factor falls below a critical one, the flickering disappears 
(since the Doppler boosting greatly weakens). Moreover, beyond this length, the jet experiences 
significant deceleration. Considering observational evidence of sub-relativistic ejecta
from nuclear regions of the Galactic center \citep[e.g.,][]{Rauch2016} and other low-luminosity
AGNs \citep{Middelberg2004}, we choose case 1) that this 
jet is slowed down at $H_{\rm p}/3$ from relativistic states to sub-relativistic ones.
The bulk Lorentz factor of the jet is similar to that of blazars
\citep{Ghisellini1993}. It should be noted that the Lorentz factor is fully independent
of the sMBH because both $L_{\rm BZ}$ and $\Delta M_{\rm j}$ are proportional to the square
of the sMBH masses. From Eq.(\ref{eq:Gamma}), we know the choked length is a significant fraction
of the thickness of the SMBH-ADAF. The appearance of the jet strongly depends on the location 
of the sMBH as the result of SMBH-ADAF density. If the sMBH is located at around $20\,\Rg$ or 
so, the jet is capable of penetrating the entire SMBH-ADAF, showing superluminal motions of
blobs. Considering the Doppler boosting effects, we have the observed luminosity of the jet inside 
the ADAF
\begin{equation}\label{eq:Ljet}
L_{\rm jet}= L_{\rm BZ}\calD^{4}
    =1.92\times 10^{36}\,\alpha_{0.1}^{-1}j_{\bullet}^{2}\calD_{10}^{4}\mathdotM_{\rm p,\bar{3}}
%    \left(\frac{m_{1}}{4}\right)^{2}\left(\frac{M_6}{4}\right)^{-1}
     q_{\bar{5}}^{2}\left(\frac{M_6}{4}\right)\,{\ergs},
\end{equation}
where $\calD_{10}=\calD/10$, $\calD=1/\Gamma(1-\beta_{\rm j}\cos\theta)$ is the Doppler factor, 
$\Gamma=1/\left(1-\beta_{\rm j}^{2}\right)^{1/2}$ is the Lorentz factor, $\beta_{\rm j}=v_{\rm j}/c$ 
is the jet velocity, and $\cos\theta$ is the cosine of the viewing angle. Here we take
$\theta=5.7^{\circ}$ and $\Gamma=8.5$ for $\calD=10$. We note that the SMBH-ADAF is optically 
thin ($\tau_{\rm es}\sim0.008\ll 1$), which allows observers to see the emissions from the 
relativistic part of the choked jets. The choked part of the jet becomes ejecta, which carries the 
rest of the BZ power and can emerge with a sub-relativistic velocity of
\begin{equation}\label{eq:veje}
\frac{v_{\rm eje}}{c}\approx \left(\frac{\xi L_{\rm BZ}\Delta t_{\rm cav}}{\Delta M_{\rm j}c^{2}}\right)^{1/2}
     =0.53\,\xi_{0.1}^{1/2}\alpha_{0.1}^{-1/2}j_{\bullet}x_{\rm j,1}^{-1}
     r_{1}\left(\frac{\ell_{\rm j}}{H_{\rm p}}\right)^{-1/2},
\end{equation}
where $\xi_{0.1}=\xi/0.1$ is the fraction of the BZ power remaining after the jet radiates its 
most part of the BZ power. This is consistent with the wisps in \SgrA observed by \cite{Rauch2016}. 
If the relativistic ejecta is entirely 
choked, the emissions could be too faint to detect. In such a case, no flickering appears.

As a result of the continual heating of sMBH-outflows, the accumulated energies give rise to a 
flare. Therefore the number of flickerings can be estimated by considering the cooling and cavity 
timescales before
a flare happens in the presence of the instability of the SMBH-ADAF (the regions with a radius 
of $R_{\bullet}\sim H_{\rm p}$ described by Eq.\ref{eq:Rflare}). It follows from
\begin{equation}
N_{\rm flick}=\frac{\Delta t_{\rm cool}}{\Delta t_{\rm cav}}
             \approx 7.4\,\left(\frac{\Delta t_{\rm cool}}{5.6\,{\rm hour}}\right)
             \left(\frac{\Delta t_{\rm cav}}{45.9\,{\rm min}}\right)^{-1}.
\end{equation}
This indicates that flares always appear after about a couple of flickerings. This is an important 
feature to observationally test the present model. Indeed, this is
consistent with observations of \SgrA \citep[see the {\it Spitzer} data of][]{Boyce2022}. 

The $L_{\rm BZ}$ (Eq.\,\ref{eq:LBZ}) is converted
into the non-thermal emissions from the relativistic jet. Internal shocks are formed due to 
successive ejecta from the sMBH-ADAF whereas external shocks are formed through the collisions 
between the jet and the SMBH-ADAF. This process is similar to GRBs \citep[e.g.,][]{Meszaros2002}.
Moreover, the jet is undergoing mass-loading processes when it penetrates through the SMBH-ADAF.
The temporal profiles of flickerings, as discussed previously, depend on both the Bondi accretion 
processes and the propagation of the relativistic jet inside the SMBH-ADAF. Details of the temporal 
profiles of the flickerings can be further understood by numerical simulations.
We would like to point out the diversity of the sMBH-driven phenomena. 
In the high-$\mathdotM$ SMBH-ADAFs, the jet is seriously choked so that the Doppler 
boosting effects are too faint to see the flickerings, however, the sMBH-outflows can still trigger
flares through viscous instability. 

\subsection{SED of the jet}
In order to calculate the SED from the relativistic jet, we have to know the magnetic fields and
energy distributions of non-thermal electrons. One-zone model is the simplest, in which non-thermal 
electrons, magnetic fields, and bulk Lorentz factor of the jet are homogenous and is often employed 
for the canonic SED of blazars \citep[e.g.,][]{Inoue1995}. In the current case of sMBH-ADAF, 
the magnetic fields are so strong (than that of blazars) that SEDs generated by this model have too 
high-frequency cutoff even in the near-infrared bands (see Figure\,\ref{fig:sed_sgr} for smaller index 
$m$). Therefore, we employ a simplified version of the inhomogeneous 
model of jets \citep{Ghisellini1985,Georganopoulos1998}. Since both geometries of the jet and its 
magnetic fields are poorly understood, we consider a simple cone of the jet and its cross-sectional 
radius increases with height ($z_{\rm j}$) as $X_{\rm j}=X_{0}\left(z_{\rm j}/{z_{0}}\right)^{m}$,
where $X_{0}$ is the jet radius at the base $z_{0}$ and index $m$ describes the geometry.
In light of the self-similar solution \citep[e.g.,][]{Narayan1995}
\begin{equation}
B_{\rm disk}=2.9\times 10^{4}\,\alpha_{0.1}^{-1}(1-\beta_{\rm s})^{1/2}
              \mathdotM_{\rm p,\bar{3}}^{1/2}x_{1}^{-5/4}\left(\frac{M_{6}}{4}\right)^{-1/2}\,{\rm G},
\end{equation}
where $\beta_{\rm s}$ is the magnetization parameter of the sMBH-ADAF, $x_{1}=X_{0}/10\,r_{\rm g}$ is 
the radius of the sMBH-ADAF. We assume that the poloidal magnetic fields of the jet (i.e., $B_{\perp}$ 
in Eq.\,\ref{eq:LLBZ}) at its base follow the sMBH-ADAF, and have 
%the magnetic fields are independent of the sMBH mass.  
%\citep[e.g.,][]{Georganopoulos1998}, we 
%obtain the magnetic fields in the jet frame as
%
\begin{equation}
B_{\rm j}=B_{\rm disk}\left(\frac{X_{\rm j}}{X_{0}}\right)^{-2}
         =B_{\rm disk}\left(\frac{z_{\rm j}}{z_{0}}\right)^{-2m},
\end{equation}
based on a simple conservation of the magnetic fluxes. In this paper, we skip the details of electron 
acceleration, which involve many processes \citep[formation and diffusions of shocks, energy gains and 
losses during the propagation of the jet through the SMBH-ADAF; see details in][]{Blandford1987}.
In the present paper, we assume a power law distribution of the electrons as 
\begin{equation}
N_{\rm e}=N_{0}\gamma^{-s}\left(\frac{z_{\rm j}}{z_{0}}\right)^{-2m}
               \quad {\rm for}\,\, \gamma_{\rm min}\le \gamma\le \gamma_{\rm max},
\end{equation}
where $\gamma$ is the Lorentz factor of non-thermal electrons, $\gamma_{\rm min,max}$ are the minimum 
and maximum, respectively, $s$ is the electron index, 
$N_{0}=(s-1)N_{\rm tot}/\gamma_{\rm min}^{1-s}
\left[1-\left(\gamma_{\rm max}/\gamma_{\rm min}\right)^{1-s}\right]$, and
$N_{\rm tot}$ is the number density of relativistic electrons. The factor $(z_{\rm j}/z_{0})^{-2m}$ 
results from the mass conservation in a continuous jet. We take $\gamma_{\rm min,max}$
as constants along the jet. When $m\rightarrow 0$, the present model
tends to the one-zone model.

Emissions from the jet in its co-moving frame can be expressed by
\begin{equation}
L_{\nu}^{\rm Syn}=\int_{z_{0}}^{\ell_{\rm j}} \pi X_{\rm j}^{2} j_{\nu}^{\rm Syn}(z_{\rm j})\,
%    \frac{j_{\nu}^{\rm syn}(z_{\rm j})}{\kappa_{\nu}(z_{\rm j})}     e^{-\kappa_{\nu}z_{\rm j}}
    dz_{\rm j};\quad {\rm and}\quad
L_{\nu}^{\rm IC}=\int_{z_{0}}^{\ell_{\rm j}}\pi X_{\rm j}^{2}j_{\nu}^{\rm IC}(z_{\rm j})\,dz_{\rm j},
\end{equation}
%\begin{equation}
%L_{\nu}^{\rm Syn}=\pi X_{\rm j}^{2}\ell_{\rm j} j_{\nu}^{\rm Syn};\quad {\rm and}\quad
%L_{\nu}^{\rm IC}=\pi X_{\rm j}^{2}\ell_{\rm j}j_{\nu}^{\rm IC},
%\end{equation}
%
where $j^{\rm syn}_{\nu}$ and $j^{\rm IC}_{\nu}$ are the synchrotron and IC 
emissivities, respectively. We then transform these into the observer's frame
in order to compare with observations \citep[e.g.,][]{Lind1985}. We neglect the synchrotron
self-absorption since most photons are radiated from the sides of the jet and are
beamed by the bulk motion of the jet, even if the optical depth is larger than unity along
the jet direction. We neglect the pair production of the IC photons.
The total number of non-thermal electrons $(N_{\rm tot})$ is constrained by
%
%\begin{equation}
$L_{\rm BZ}=\int_{z_{0}}^{\ell_{\rm j}}\pi X_{j}^{2} dz_{\rm j}
            \int_{\nu_{1}}^{\nu_{2}}\left(j^{\rm Syn}_{\nu}+j^{\rm IC}_{\nu}\right) d\nu$,
where $\nu_{1}$ and $\nu_{2}$ are frequencies of the synchrotron radiation and IC. 
We use the standard formulations of synchrotron self-Compton (SSC) emissions and
IC \cite[]{Blumental1970} for the simple SED of the jet. 
%The energy density inside the SMBH-ADAF could be a complicated function of space for external
%inverse Compton scattering. 
We approximated it an isotropic radiation field. The ADAF is taken 
to be a point source with a luminosity $L_{\rm ADAF}$ and an averaged energy density of 
$u_{\rm ADAF}=L_{\rm ADAF}/4\pi R^{2}c$ at a distance $R$. In the co-moving frame, the 
relativistic jet receives an energy density of the SMBH-ADAF given by
$U_{\rm ADAF}\approx \Gamma^{2}u_{\rm ADAF}$,
which is the external source of seed photons \citep[][]{Sikora1994}. We ignore external Compton 
scattering since $U_{\rm B}/U_{\rm ADAF}\sim 10^{6}$ holds in the current parameters of the jet.
We omit all the formulations in this paper. Generally, there are two peaks of SEDs 
arising from synchrotron and IC, respectively.

\begin{figure}
    \centering
   \includegraphics[width=0.9\textwidth,trim=15 0 30 0,clip]{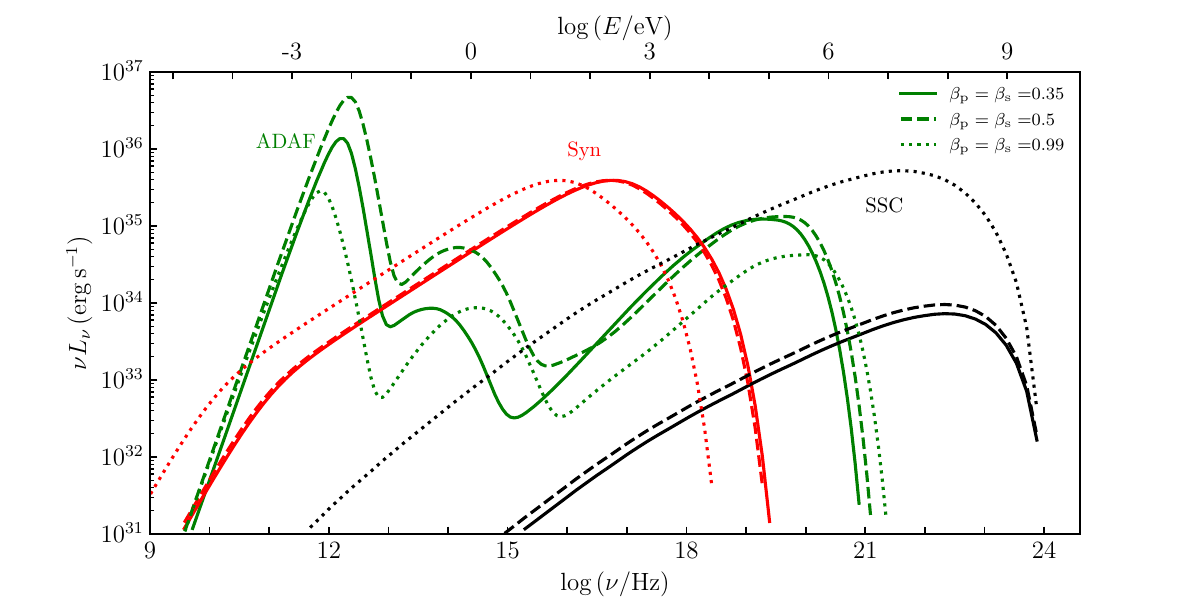}
    \caption{\footnotesize The overall spectral energy distributions of the SMBH-ADAF 
    and the relativistic jets from the sMBH. The relativistic jet mainly contributes to the optical 
    to soft X-ray bands. In these calculations, we take the same values of the magnetization parameter 
    ($\beta_{\rm p,s}$) in the SMBH- and sMBH-ADAFs. See text for values of model parameters.}
\label{fig:sed_mod}     
\end{figure}

We take $M_{6}=4.0$, $\mathdotM_{\rm p,\bar{3}}=1$, $\alpha_{0.1}=1$. Following 
\cite{Narayan1995} and \cite{Manmoto2000}, we take the electron heating coefficient $\delta=0.03$ 
in this paper though it may have large uncertainties \citep[e.g.,][]{Yuan2003}.
We calculate the SMBH-ADAF SED for $\beta_{\rm p}=(0.99,0.5,0.35)$.  
Figure\,\ref{fig:sed_mod} shows the results. Generally, SEDs of the SMBH-ADAF are consistent 
with those of \cite{Manmoto2000}. The first peak of the SMBH-ADAF arises from synchrotron radiation
of the Maxwellian distributions of hot electrons, and the peak shifts  
and powers vary with $\beta_{\rm p}$. Comptonization of the hot electrons results in the
second peaks and bremsstrahlung contributes to the last peak with a cutoff related to
the maximum temperatures of electrons. 

For parameters of the jet, we take $x_{\rm j,1}=1$, $\ell_{\rm j}=H_{\rm p}/3$, $\Gamma=8.5$ 
and $\theta=5.6^{\circ}$. We fix the jet location at the radius of $x_{1}=1$. 
For a jet with $m=1/5$, the top of the jet
$z_{\rm j, top}=\ell_{\rm j}\approx 10^{12}\,$cm, $\ell_{\rm j}/z_{0}\sim 10^{4}$,
we have $X_{\rm j}\approx 7\,X_{0}$ and the magnetic fields decay by a factor of 50 along the 
jet height. The free parameters are $\gamma_{\rm min}$, $\gamma_{\rm max}$, $s$, $m$ and 
$\beta_{\rm s}$. We take $s=2.0$, $\gamma_{\rm min}=1$ and $\gamma_{\rm max}=10^{3}$ and 
$m=1/5$ for the theoretical SEDs. We adjust
$\beta_{\rm s}$ in order to show the role of the magnetic fields in the jet SED.
Since we keep the power of jets as a constant ($L_{\rm BZ}$), 
we adjust electron numbers for the dependence of the SED on magnetic fields, $\gamma_{\rm min,max}$,
$\beta_{\rm s}$, $s$ and $m$. The synchrotron emissions shift towards lower frequencies with increases
of $\beta_{\rm s}$ because $B$ decreases. As shown in Figure\,\ref{fig:sed_mod},
the synchrotron emissions from the jet just supplement the deficits of the SEDs from SMBH-ADAFs 
depending on the maximum energies of electrons ($\gamma_{\rm max}$), in particular, 
contributing to infrared to soft X-ray bands. The SED of the jet shifts toward low frequency with 
decreases of $m$. This results from the fact that magnetic fields decrease along jet height with $m$.
When $m$ tends to zero, the jet tends to the one-zone model.

IC of synchrotron photons is significant compared with the power of the synchrotron emissions. 
It strongly depends on the magnetic fields of the sMBH-ADAF. In $\beta_{\rm p,s}=0.99$ case 
(weak magnetic field) as shown in Figure\,\ref{fig:sed_mod}, the self-inverse Compton power 
is comparable with that of the synchrotron radiation. This significantly contributes to 
$\gamma$-ray bands but relies on $\gamma_{\rm max}$. With increases of magnetic fields 
($\beta_{\rm s}$ decreases), the IC power decreases dramatically. 

For a simple treatment, we take a constant bulk Lorentz factor of the jet in the present model. 
Dynamical interaction with the SMBH-ADAF will slow down the jet. A self-consistent treatment
of this is necessary \citep[e.g.,][]{Contopoulos1995} since the fully parameterized model of 
inhomogeneous jets could be useful to explore the SEDs of the sMBH-driven jets like in blazars 
\citep{Georganopoulos1998}. Moreover, it is necessary to explore 
the time-dependent model of the relativistic jets for the temporal profiles of the light curves.
This is left for future research. We would like to point out that the BZ process can also occur 
in the SMBH-ADAF if the SMBH is rotating fast enough. In such a context, 
emissions of the sMBH-ADAFs are overwhelmed by the SMBH-ADAFs. BL Lac objects and face-on radio 
galaxies (Fannaroff-Rilley I radio galaxies) are known to contain ADAFs which power relativistic 
jets \citep{Sikora2007,Cao2004,Tadhunter2016}. Actually, LLAGNs often show large radio-loudness 
\citep[e.g.,][]{Ho2002} but lack powerful jets \citep[e.g.,][]{Middelberg2004}, implying 
that the SMBHs in most LLAGNs are non-rotating. There is
no evidence that the BZ process works in SMBH-ADAF of \SgrA, which only shows sub-relativistic 
wisps discovered by \cite{Rauch2016}. If we apply Eq.\,(\ref{eq:LBZ}) to \SgrA, the BZ power will 
be around $10^{40}\,\ergs$ if $j_{\bullet}\sim 1$, which is much more luminous than the observed
the total emissions of Sgr A$^{\!*}$.
The central SMBH is thus expected to be very slowly spinning in \SgrA. This is favored by 
the presence of two misaligned young stellar disks in light of the Lense–Thirring effects 
\citep{Fragione2022}, though Event-Horizon-Telescope observations favor 
$j_{\bullet}\gtrsim 0.5$ \citep{EHT2022}. Moreover, the observational fact that there are
two counter-rotating young stellar disks within $30^{\prime\prime}$ regions (about 1\,parsec)
of the Galactic center
identified by SINFONI IFU at the VLT \citep{vonFellenberg2022} directly indicates random
accretion onto the central SMBH, making the SMBH spin very low \citep{Wang2009,Volonteri2013}. 
Independent evidence for random accretion onto the central SMBH is provided by ALMA
that two counter-rotating disks of gas in NGC\,1068 have been found by \cite{Impellizzeri2019}.
Cancellations of the angular momentum of gas ($\lesssim 10\,$pc nuclear regions) finally drive 
extremely high accretion rates 
of the SMBH, which likely leads to an super-Eddington growth of the SMBH.

\subsection{Gravitational waves}
In this paper, an sMBH orbiting the central SMBH is an excellent EMRI (extreme mass ratio inspiral). 
We assume that the EMRI follows a circular orbit. This can be justified by the decaying of 
ellipticity due to radiations of
gravitational waves (GWs). According to \cite{Peters1964}, we have the circularization 
timescale $t_{\rm E}=\left(d\ln e/dt\right)^{-1}$, where $e$ is the ellipticity of the initial
orbit, if the GWs radiations govern the evolution of the EMRI orbit. A highly elliptical 
orbit of an EMRI will be circularized with 
$t_{\rm E}=\left(d\ln A/d\ln e\right)t_{\rm GW}\approx (1-10)\,t_{\rm GW}$ 
from an initial elliptical orbit of $e_{0}=0.9$ from $A=100\,\Rg$, where $A$ is the separation of
the EMRI, 
$d\ln A/d\ln e=12\left[1+(73/24)e^{2}+(37/96)e^{4}\right]/19(1-e^{2})\left[1+(121/304)e^{2}\right]$. 
We expect that the sMBH undergoes rapid circularization of orbits and reaches a circular orbit at 
$A=10\Rg$. Their strain amplitudes and frequency of GWs from the EMRI are given by
\begin{equation}\label{eq:hs}
h_{\rm s}=\left(\frac{128}{15}\right)^{1/2}\frac{\left(GM_{\rm c}\right)^{5/3}}{c^{4}d_{\rm L}}\left(\pi f\right)^{2/3}
         =7.6\times 10^{-17}\,d_{\rm 10kpc}^{-1}a_{1}^{-1}q_{\bar{5}}\left(\frac{M_{6}}{4}\right),
\end{equation}
where $M_{\rm c}=\left(M_{\rm p}M_{\rm s}\right)^{3/5}\left(M_{\rm p}+M_{\rm s}\right)^{-1/5}
=q^{3/5}(1+q)^{-1/5}M_{\rm p}$ is the chirping mass, and
\begin{equation}
f=\frac{2}{P_{\rm orb}}=0.5\,(1+q)^{-1/2}a_{1}^{-3/2}\left(\frac{M_{6}}{4}\right)^{-1}\,{\rm mHz},
\end{equation}
where $P_{\rm orb}$ is the orbital periods, $d_{\rm 10kpc}=d_{\rm L}/10\,{\rm kpc}$ is the distance 
to observers, $a_{1}=A/10\Rg$ and $P_{\rm orb}$ is its orbital period. The GWs are in the bands of 
{\it LISA} (Laser Interferometer Space Antenna), and the other two space missions of {\it Taiji}
\citep{Hu2017} and {\it Tianqin} \citep{Luo2016}, whose thresholds ($h_{\rm s}\approx 10^{-20}$) are 
much lower than strains of the present EMRI. The decaying timescale of the circular orbit 
is \citep[e.g.,][]{Peters1964}
\begin{equation}
t_{\rm GW}=\frac{A}{dA/dt}
     =\frac{5a^{4}}{64q(1+q)}\left(\frac{\Rg}{c}\right)
     \approx 48.8\,a_{1}^{4}q_{\bar{5}}^{-1}\left(\frac{M_{6}}{4}\right)\,{\rm yr},
\end{equation}
which is a feasible timescale to witness an EMRI merger. We note that this 
timescale is very sensitive to the separation of the EMRI, and expect GRAVITY/VLTI to
make a precise measurement of the orbit from the flares in \SgrA.

In summary, an sMBH residing in the ADAF of the central SMBH has to undergo an episodic Bondi 
accretion governed by its strong feedback. A cavity is formed by the outflows from the accretion. 
During the accretion, a relativistic jet is formed by the BZ mechanism, showing quasi-periodic 
flickerings (jet emissions) from the sMBH-ADAF. Accumulations of the outflow energies will trigger
the viscous instability of the SMBH-ADAF and generate a flare subsequently. Milli-Hz gravitational 
waves are radiated by the EMRI, which is strong for the detection of the designed space missions.
Table\,\ref{tab:model_para} lists all the parameters involved in this model. Given a sMBH-SMBH system 
in the ADAF state, the temporal properties of the system can be predicted for quasi-periodic flickerings 
and flares.

\begin{deluxetable*}{lll}
\tablecaption{Parameters of the sMBH-SMBH system (EMRI)\label{tab:model_para}}
\setlength{\tabcolsep}{1pt}
\tabletypesize{\footnotesize}
\tablehead{Parameters &$\quad\quad$ & Meanings }
\startdata
\multicolumn{3}{l}{Parameters describing the system given by the initial conditions}\\ \hline
$M_{\rm p,s}$ & &Masses of the SMBH and sMBH, respectively; the mass ratio is defined by $q=M_{\rm s}/M_{\rm p}$  \\
$\dot{\mathscr{M}}_{\rm p,s}$ & & Dimensionless accretion rates of the SMBH and sMBH in units of 
$L_{\rm Edd}/c^{2}$, respectively\\
$R_{\bullet}$         & &sMBH location radius at the SMBH-ADAF (or $A$: the separation between sMBH and SMBH) \\ \hline
\multicolumn{3}{l}{Parameters of accretion physics}\\ \hline
$\alpha_{\rm p,s}$    & &Viscous parameters of the SMBH and sMBH\\
$\beta_{\rm p,s}$     & &Magnetization parameters of the SMBH and sMBH\\
$\eta_{\rm out}$      & &Efficiency of outflows driven by sMBH-ADAF\\ \hline
\multicolumn{3}{l}{Derived parameters of the system for observations of flickerings and flares}\\ \hline
$R_{\rm cav}$         & &Radius of the cavity created by the SMBH-ADAF outflows \\
$\Delta t_{\rm cav}$  & &Formation timescale of the cavity as the quasi-periods of flickerings \\
$\Delta t_{\rm cool}$ & &Cooling timescale of the SMBH-ADAF as outburst timescales of flares\\
$R_{\rm flare}$       & &Radius of the cavity zones generating flares from the SMBH-ADAF\\
$L_{\rm BZ}$          & &Blandford-Znajek power of the sMBH     \\
$j_{\bullet}$         & &Specific angular momentum of the sMBH  \\
$\Gamma$              & &Lorentz factor of the relativistic jet developed by the sMBH-ADAF\\
$v_{\rm eje}$         & &Sub-relativistic velocity of the choked part of the relativistic jet\\
$N_{\rm flick}$       & &Numbers of flickerings triggering a flare \\ 
%$A$                   & &Separation between the sMBH and SMBH \\
$h_{\rm s}$           & &Strains of the gravitational waves from the sMBH-SMBH binary system\\
$f$                   & &Frequency of the gravitational waves\\
$t_{\rm GW}$          & &Timescales of orbital decays due to gravitational waves\\ \hline
\multicolumn{3}{l}{Parameters of the relativistic jet (assumed in this paper)}\\ \hline
$X_{\rm j}$           & &Cross-sectiobal radius of the jet; $X_{0}$ is the initial radius \\
$\ell_{\rm j}$        & &Length of the jet; $z_{0}$ is the initial height of the jet; $z_{\rm j}$ is jet height\\
$s$                   & &Distribution index of non-thermal electrons of the relativistic jet ($\propto \gamma^{-s}$)\\
$m$                   & &Geometric index of the cross-sectional radius of the jet versus its height ($X_{\rm j}\propto z_{\rm j}^{m}$)\\
$\gamma_{\rm min,max}$& &Minimum and maximum Lorentz factors of non-thermal electrons\\
\hline
\enddata
%\tablecomments{The present discussions of the paper are valid for a central SMBH with ADAFs and
%$q\ll 1$.
%}
\end{deluxetable*}

\subsection{Discussions}
In this paper, the AMS plays a role in perturbations of the SMBH-ADAF while the latter is in a
relatively stationary state. Flickerings are a tiny fraction of the SMBH-ADAF radiation but the 
flares are stronger than the former. Therefore, light curves are expected to show as depicted by 
the right panel of Figure\,\ref{fig:model}, where flares and flickerings are thus just superposed 
on a relatively stationary radiation flux. If the sMBH is non-rotating (or its rotation is not 
fast enough), flickerings disappear (the lower panel). We denote this type-A light curve, and
type-A1 and type-A2 for the cases with and without flickerings, respectively. Actually, \SgrA 
shows the type-A light curves, see Figure 1 in \cite{Boyce2022}, NGC\,4151 
\citep[see Figure 4 in][]{ChenYJ2023}, and NGC\,5548 \citep[see Figure 3 of][]{Li2016}.

For high-$q$ system, however, feedback of the sMBH accretion to the SMBH-ADAF is not a perturbation. 
Large zones of the SMBH-ADAF will be broken, and giant flares are expected from these high-$q$ systems.
We hence anticipate an upon-down style of light curves. This denotes type-B light curves. Similar to 
type-A curves, type-B curves are distinguished as type-B1 and type-B2 with and without flickerings, 
respectively. Actually, Arakrian\,120 has exhibited type-B light curves over the last 20 years \citep{Li2019}.
Collections of low-$\mathdotM$ AGN light curves can test this classification,  but it needs homogeneous 
light curves spanning longer than 20-30\,years.
We emphasize that these classifications of light curves are only valid for the case
that the central SMBHs have ADAFs, and the physics for SMBHs with high-$\mathdotM$ should
be reconsidered separately.

In the present study, we set the typical values of an AMS with a mass of $\sim 40\sunm$ at 
$10\,\Rg$ around the central
SMBHs of $4\times 10^{6}\sunm$ for the Galactic Center. The resultant properties of the AMS
depend on its mass, location and the SMBH-ADAF accretion rates. The BZ power is proportional 
to $M_{\rm s}^{2}$ (see Eq.\,\ref{eq:LBZ}). Fates of the relativistic jet depend on the
SMBH-ADAF. It either penetrates the ADAF and shows superluminal motion outside the nucleus, 
or is partially choked by the ADAF giving rise to sub-relativistic ejecta from the nucleus
(see Eq.\ref{eq:veje}). If the BZ power is not strong enough compared with the SMBH-ADAF damp, 
flares still occur but without flickerings. Moreover, the sMBH has a Keplerian rotation
velocity of $v_{\rm rot}\sim c/3$ at $R_{\bullet}\sim 9\,\Rg$, emissions from the jet developed by the 
sMBH-ADAF could be modulated by transverse Doppler boosting. Additionally, general relativistic 
effects should be employed for temporal profiles of flickerings. 

Shocks formed by the dynamic interaction between sMBH-outflows and SMBH-ADAF can accelerate electrons 
and generate non-thermal emissions. As a simple estimation, the kinetic power of the shocks is around
$L_{\rm out}\approx 8\times 10^{33}\,\ergs$, and about 10\% of $L_{\rm out}$ will be channelled into
non-thermal electrons \citep[e.g.,][]{Blandford1987} and contribute to multiwavelength continuum.
%As shown subsequently in \S\ref{sec:flickering}, this population of electrons
%could explain the radio continuum ($\le 100\,$GHz) of \SgrA as an extra component from the
%pure ADAF emissions \citep{Markoff2001,Yuan2003}. 
This component causes complicated behaviors 
of variabilities of the system. We remain this topic as a future issue. 

We assume that the global SMBH-ADAF is stationary. It is then expected to exhibit quasi-periodicity 
of flickerings and flares from the system. However, the reality could be complicated. For example, 
$\mathdotM_{\rm p}$ is a function of the radius of SMBH-ADAF \citep[][]{Blandford1999}, and due to
clump accretion \citep{Wang2012a}, the periodicity of flickerings and flares dramatically decreases
and even becomes random sometimes. For example, NGC\,5548 shows preliminary periodicity in its long-term 
light curves with flickerings \citep{Li2016}, but the periodicity of flickerings is usually verily 
twinkled by showing 3-4 cycles.

Finally, we would like to point out the key tests of the present model. LISA  
detections around 2030 can more accurately determine the EMRI
system for a concluding remark. Before the LISA era, GRAVITY+/VLTI observations 
of \SgrA can provide more solid and accurate loci of NIR-flares from increasing observations
to obtain the orbiting radius of flares, on which the merger timescale sensitively 
depends. The current error bars of the flare's loci are still quite large (about $30-50\%$). 
Flickering numbers associated with flares are another feature of the EMRI system, and we 
anticipate acquiring more data for high statistics of the flares and flickerings.

\section{Application to \SgrA}\label{sec:application}
As a preliminary practice of the present model, we apply it to \SgrA, which is the best-studied 
low-luminosity system. An early extensive review on this object can be found for general properties
in \cite{Genzel2010}. It has been extensively observed through multiwavelength 
campaigns during the last 20\,years and shows a diversity of variability properties 
\citep[see a brief summary of observations in][]{Witzel2021,Boyce2022, vonFellenberg2023}. 
The powerful astrometric
measurements of GRAVITY/VLTI provide an exciting ring of locations of flares during the last 
four years. The ring is phenomenologically explained as a moving hot spot around the
SMBH. It has a radius of $R_{\bullet}=8.9_{-1.3}^{+1.5}\,\Rg$ 
and is rotating with the azimuthal speed of near Keplerian motion
\citep[see also their Figure 7 in][]{Gravity2023}. A hot spot due to magnetic reconnection in 
the SMBH-ADAF has been suggested for flares in \SgrA (see references subsequently cited), but 
an orbiting sMBH can explain the 
properties of flares and flickerings as an alternative model. Actually, the light curves show 
that flickerings and flares are superposed on relatively stationary fluxes 
\citep[see {\it Spitzer, Chandra} light curves][]{Boyce2022}, implying that they are
perturbations of the SMBH-ADAF. This indicates that \SgrA should be classified as one type-A2
object.

\begin{figure}
    \centering
    \includegraphics[width=0.95\textwidth,trim=25 0 30 0,clip]{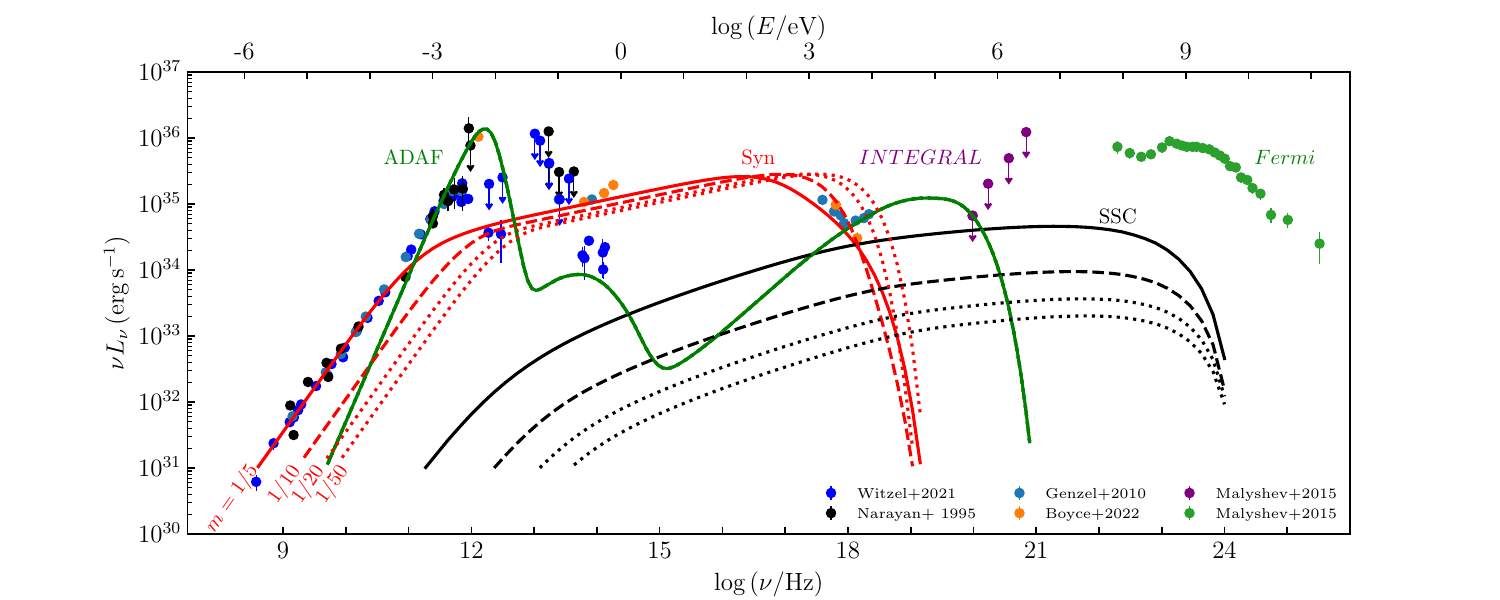}
    \caption{\footnotesize The overall spectral energy distribution of \SgrA from radio to 
    TeV bands. We get the accretion rates of the SMBH from the averaged SED.  Curves are marked by 
    the same color words. Syn: synchrotron radiation from the relativistic jet developed by the sMBH 
    accretion. ADAF-EC: external IC of SMBH-ADAF photons. SSC: 
    synchrotron self-Comptn scattering. \cite{Narayan1995} collected early radio data (we omit these 
    references). {\it INTEGRAL} and {\it Fermi} data are from \cite{Malyshev2015}. When $m\rightarrow 0$,
    the inhomogeneous model becomes one-zone model.
%     The $\gamma$-rays in the {\it MAGIC} bands could be 
%    ascribed to unknown objects nearby \SgrA \citep{Ahnen2017,Acciari2020}.
     }
\label{fig:sed_sgr}     
\end{figure}

\subsection{Accretion rates}
%
%\SgrA is the best-studied low-luminosity SMBH accretion system with radiations
%spanning from radio to $\gamma$-rays. 
 The SMBH mass is 
accurately measured $M_{\rm p}=4.3\times 10^{6}\,\sunm$ 
\citep[see the latest values given by][]{Gravity2022}. The classical ADAF model was first 
applied to explain SED (from radio to hard X-rays) of 
\SgrA \citep{Narayan1995b}, which was revised by including fully general relativistic effects
\citep[e.g.,][]{Manmoto2000,Li2009}. Figure \ref{fig:sed_sgr} shows the global SED from radio to 
TeV bands. We use the classical ADAF model \citep[e.g.,][]{Li2009} to get the accretion rates 
for discussions on the sMBH properties. $\gamma$-ray emissions through hot proton-proton 
collisions in the ADAFs have been suggested by \cite{Mahadevan2003}, but 
it is only a small fraction of the total ADAF emissions \citep{Oka2003} that is much below
the {\it Fermi} observations as shown by Figure\,\ref{fig:sed_sgr}.
For \SgrA, there are only three free parameters in this model, accretion rates 
($\mathdotM$), viscosity ($\alpha$) and magnetization parameter ($\beta_{\rm p}$).
We take the SMBH mass measured by GRAVITY \citep{Gravity2022}, and fix the typical values 
of the magnetization parameter of $\beta=0.35$ and the viscosity of $\alpha=0.1$. We find 
$\mathdotM_{\rm p}=10^{-3}$ for a global fitting (the green line for most points 
except for flaring points in Figure\,\ref{fig:sed_sgr}), which agrees with previous results of
\cite{Manmoto2000}.

It has been also suggested that the X-ray flare emission is due to synchrotron radiation 
\citep{Dodds-Eden2009,Barriere2014,Ponti2017} although it has been also interpreted as 
IC upscattered photons by the mildly relativistic, nonthermal electrons 
\citep{Markoff2001,Yuan2003,Yusef-Zadeh2009,Ball2016}. It is obvious that the {\it Fermi} 
$\gamma$-ray emissions \citep[][]{Chernyakova2011,Malyshev2015} are much beyond the scope 
of the SMBH-ADAF continuum emissions, but they have a luminosity of $10^{36}\,\ergs$ from 
100\,MeV to 500\,GeV. It has been suggested by \cite{Malyshev2015} that the {\it Fermi}-detected
bump originates from IC by high-energy electrons in extensive regions, indicating 
an extra source of non-thermal emissions in \SgrA 
\citep[see the possibilities discussed in][]{Cafardo2021}. 
%In this paper, we suggest that this component originates from the relativistic jet developed 
%from the sMBH-ADAF as described below. 
It is not our goal to extensively explore the delays among the multiwavelength variations, 
but we explain the major properties of flickerings and flares and suggest future tests of the
present model.

\subsection{Flares and quasi-periodic flickerings}\label{sec:flickering}
NIR flares are more common on timescales of hours or so. X-ray flares often accompany NIR ones, but 
sometimes don't \citep{Boyce2022}. Several kinds of models are suggested to explain multiwavelength 
variations, such as, the popular magnetic reconnection events \citep[e.g.,][]{Dexter2020,Mellah2023},
non-thermal electrons in a jet \citep{Markoff2001,Yuan2003}, sudden instabilities of the 
MHD disk \cite[e.g.,][]{Chan2009}, or other 
stochastic processes in the ADAFs \citep[see more references listed by][]{Boyce2022}. The most 
interesting is that \cite{Genzel2003} discovered quasi-periodic flickerings of $\sim 17\,$min, which
were superimposed to a flare in NIR bands during one epoch of 2002 (but see different results in other 
epochs of \citealt{Do2009}). X-rays of {\it XMM-Newton} observations show the similar 
periods \citep{Aschenbach2004,Eckart2006}. This quasi-periodic flickering (at 4$\mu$m) has been 
confirmed by the {\it Spitzer} observations \citep{Boyce2022}. It has been argued that the appearance 
of the flickering quasi-periodicity ($17\sim40\,$min) depends on epochs \citep[see a summary of 
the quasi-periodicity in][]{Genzel2010} though \cite{Do2009} thought of the res-noise roles. The 
quasi-periodicity of flickerings has been suggested to arise from the quasi-periodic structure of 
plasma as hot spots \citep[e.g.,][]{Dexter2020,Mellah2023,Aimar2023,Lin2023} 
or expanding hot spot \citep{Michail2023}. However, this kind of models involving magnetic
reconnections should explain why the flares constantly happen at the same radius (namely, $\sim 9\Rg$) 
since the reconnections randomly happen somewhere inside the SMBH-ADAF.
Second, one orbiting star as a pacemaker is suggested by \cite{Leibowitz2021} for the periodicity 
of NIR and X-ray flares. A hidden black hole with a mass of $\sim 10^{4}\sunm$ has been suggested 
by \cite{Naoz2020}, however, the black hole with $\gtrsim 10^{3}\sunm$ has been ruled out by precise 
measurements of S2 orbits \citep{Gravity2023b,Will2023}. Recently, \cite{Gravity2023} maps the loci 
of the flare's locations during the last 4 years and finds that the loci are consistent with the 
Keplerian orbits at a distance of $\sim 9\,\Rg$ from the central SMBH well determined by
the S2 star. For the magnetic reconnection model of the hot spots, the question of how to keep them 
in similar loci remains open. This new solid evidence supports a rigid body around the central SMBH. 
We suggest here an sMBH is orbiting around the SMBH in \SgrA in charge of flares and quasi-periodic 
flickering.

For simplicity, we assume the sMBH spin axis to be perpendicular to the SMBH-ADAF
equatorial plane. Since the quasi-periods are about $17-40$\,min (we take the mean quasi-period is 
bout 30\,min), we have the accretion rates of $\mathdotM_{\rm s,\bar{6}}\approx 2.0$ from 
Eq.\,(\ref{eq:Bondirate}), $\mathdotM_{\rm p,\bar{3}}=1.0$, $r_{1}=0.9$, and $M_{6}\approx 4.3$. 
We obtain the quasi-periods of $\Delta t_{\rm cav}\approx 42.1\,\alpha_{0.1}^{-1}$min 
from Eq.\,(\ref{eq:Deltatcav}), which is in agreement with observed quasi-periods \citep{Genzel2010} 
if $\alpha_{0.1}\approx 1.5\sim 2$ slightly fluctuates. Flares are generated with 
$\Delta t_{\rm vis}\approx 10.1\,$hour, which is well consistent with the observed, namely, there
are a few flares per day. The SED of the jet is shown in Figure\,\ref{fig:sed_sgr}, where we take
$s=2.6$, $\beta_{\rm s}=0.35$, $\gamma_{\rm min}=2.0$, 
$\gamma_{\rm max}=10^{3}$, $m=1/5,\,1/10,\,1/20,\,1/50$, respectively, for the 
1-100\,GHz bands, NIR bands and soft X-rays. The most significant effects are observed in the 
SED at low frequencies. The well-known 1-100\,GHz emissions beyond the SMBH-ADAF, which
are usually explained by a relativistic jet from the SMBH-ADAF lacking 
of choke effects \citep{Markoff2001,Yuan2003, 
Yusef-Zadeh2009,Ball2016}, can be reasonably explained by the present model. Ejecta from \SgrA
has been observed but it is sub-relativistic \citep{Rauch2016}. The present model is consistent 
with the ejecta. It should be mentioned that some parameters degenerate in fitting
SED, for example, $\gamma_{\rm min}$, $m$ and $\beta_{\rm s}$. The most important thing is to 
identify this EMRI system at this stage, and we leave it for the future to make a self-consistent 
model for jet emissions. However, the {\it Fermi} SED cannot be explained by the present jet model. 
We agree that {\it Fermi} $\gamma$-ray emissions are from extensive regions since no evidence 
is found for $\gamma$-ray variability \citep[e.g.,][]{Cafardo2021}.

The simple model can, in principle, explain the major properties of the flares and quasi-periodic 
flickerings. We note that the extensive multiwavelength observations show complicated behaviors 
of variations, such as delays among radio, NIR and X-ray bands. Though some attempts have been
made \cite[e.g.,][]{Okuda2023}, we will apply the present model to \SgrA in more sophisticated 
treatments to explain variabilities and polarizations by including inner shocks and external shocks
of the jet. Moreover, wisps with sub-relativistic velocity ($\sim 0.4\,$c) from \SgrA have been 
discovered by \cite{Rauch2016}, which could be the remnant, i.e., $\xi L_{\rm BZ}$, of the 
relativistic jets slowed down by interaction with surrounding medium from the sMBH-ADAF. Indeed,
LLAGs often show sub-relativistic ejecta \citep[e.g.,][]{Middelberg2004}. A damped 
jet model will be discussed for this issue in a separate paper.
A few points would be stressed as follows.

First, the non-thermal emissions from the relativistic jet can be, in practice, tested by a 
multiwavelength campaign of simultaneously monitoring Sgr\,A$\!^{*}$. In particular, the GHz 
radio and soft X-rays correlations (synchrotron emissions), and with Fermi $\gamma$-ray bands 
(external IC) are the keys to testing the present model of the sMBH. Hard X-rays may be mainly 
contributed by the SMBH-ADAF. Second, the emissions from the shocks formed by the outflows and 
SMBH-ADAF also contribute to emissions
in some bands from radio to X-rays (even soft $\gamma$-rays). This makes the tests not so direct. \cite{Okuda2023}
made an extensive analysis of the multiwavelength correlations, but they draw conclusions that the
multiwavelength continuum has complicated originations. The correlations may depend on the states of
Sgr\,A$\!^{*}$.
Third, optical and UV bands are the key bands of testing SMBH-ADAF model, but it is impossible to
have observations owing to the heavy extinctions of the Galactic center. Fortunately, {\it INTEGRAL}
(The International $\gamma$-ray Laboratory: 15\,keV-10\,MeV)
has a sensitivity ($\sim 2.85\times 10^{-6}\,{\rm ph\,s^{-1}\,cm^{-2}}$ with  an exposure time of
$10^{6}\,$s at 
100\,keV, see {\tt https://www.cosmos.esa.int/web/integral/instruments-ibis}) higher than the IC
scattering of SMBH-ADAF photons, and is promising to detect the emissions to test  
the present model. However, {\it INTEGRAL}
is not able to spatially resolve the region of the Galactic Center so that only upper limits are given
by \cite{Malyshev2015}. As to {\it Fermi} observations, the present model is not able to produce
enough $\gamma$-ray emissions to explain the data. We agree that these $\gamma$-rays are from
extended regions or other compact objects in the Galactic Center \citep{Cafardo2021}.

\subsection{mHz gravitational waves}
If the $40\sunm$ black hole is orbiting around the central SMBH, \SgrA will be an excellent
target of LISA detection \citep{Babak2017}. From Eq.\,(\ref{eq:hs}), we find the strain amplitude 
$h_{\rm s}=1.4\times 10^{-17}$, 
which is very strong for LISA. The decaying timescale of the orbits is 
$t_{\rm GW}\approx 32.9_{-15.4}^{+28.4}\,$years for $q_{\bar{5}}=1$, $M_{6}=4.3$ and $a_{1}=0.9$ 
in light of the current error bars of the orbital radius 
($A$) from GRAVITY. More data are needed from GRAVITY to determine $A$ better with higher
statistics. The key test is from LISA to detect the strains and polarizations of the mHz 
gravitational waves from the EMRI and its orbital decays. See some detailed calculations of mHz 
gravitational waves from an EMRI, which can be found in \cite{Fang2019}, \cite{Bondani2022} and 
\cite{Tahura2022}. Considering the general relativistic effects, the Schartzshild procession of 
the sMBH is about $\Delta \phi\sim 6\pi \Rg/R_{\bullet}\approx 108^{\circ}\,r_{1}^{-1}$ per period 
for a circular orbit. Therefore, it is feasible for GRVAVITY+/VLTI to measure the procession and 
decay of the orbit (through loci of the flare's location). Fully general 
relativistic treatments should be simultaneously done for both Schwartzschild procession and
gravitational waves. \SgrA is an excellent laboratory for general relativity through gravitational 
waves detected by {\it LISA\,/\,Taiji\,/\,Tianqin}, and orbits measured by GRAVITY+/VLTI.

\section{Conclusions}
We outline a model of the case of a stellar-mass black hole (sMBH) as one satellite of the SMBH 
embedded in the advection-dominated
accretion flows (ADAFs). The Bondi accretion onto the sMBH drives the formation of a cavity through
outflows and leads to quenching the accretion. A cavity is expected to quasi-periodically appear in the 
SMBH-ADAF, and accumulated energies of sMBH outflows during its growth will make flares through viscous
instability. Relativistic jets are developed by the Blandford-Znajek mechanism if the sMBH is maximally 
rotating, and will significantly emit non-thermal radiations spanning from radio to $\gamma$-rays. The 
non-thermal emissions from the relativistic jet follow the episodic cavities as quasi-periodic flickerings. 
Such an Extreme Mass Ratio Inspiral (EMRI) is an excellent laboratory for milli-Hertz gravitational waves.

As a simple application of the present model, we explain the flares and quasi-periodic flickerings
of \SgrA within the framework of the present scenario. GRAVITY/VLTI maps of the locations of flares 
in \SgrA consist of a ring, which supports the present model. The quasi-periodic flickerings are 
consistent with the flare's location, and flares take place driven by accumulations of about 10 
flickerings. The satellite black hole with $M_{\rm s}\approx 40\sunm$ is favored from 
fitting the SED of \SgrA spanning from radio to X-ray bands, where 
the relativistic jet is developed from the episodic sMBH-ADAF. The strain amplitudes of the mHz 
gravitational waves are about $10^{-17}$ and the sMBH will merge into the central SMBH in 30 years.
More precise measurements of the sMBH orbits are expected as well as mHz GW detections of 
{\it LISA\,/\,Taiji\,/\,Tianqin} to reveal the presence of the sMBH.

Though we only provide formulations for an EMRI with a mass ratio of $10^{-5}$ around 
$10^{6}\,\sunm$ SMBH, the present model is also applicable to $\gtrsim10^{6}\,\sunm$ system. 
Applications of the present scenario to other massive AGNs \citep[e.g.,][]{Pyatunina2006} or 
3C 390.3-like radio galaxies \citep[]{Sergeev2020} will be carried out. For high-$q$ systems, 
such as $q\sim 10^{-2}$, the properties of the system will be different from the present
descriptions, showing much larger cavities than the low-$q$ ones. In such a context, 
the EMRI system shows large amplitudes of flares 
i.e., an upon-down mode of variabilities, and flickerings appear/disappear depending on the sMBH spins. 

Finally, we would like to point out the possibility that multiple sMBHs may simultaneously  
coexist and randomly distribute inside the SMBH-ADAF. Since the properties of sMBH cavities 
are sensitive to the locations and masses, flickerings and flares are superposed on each other.
The quasi-periodicity of light curves arising from all of them 
may disappear but show random behaviors. In general, light curves of LLAGNs should be complicated
if they contain multiple sMBHs.

\acknowledgments
The authors thank an anonymous referee for a helpful report.
JMW thanks F. Yuan and Z.-Q. Shen for useful discussions as to the physics of the Galactic Center. 
Helpful discussions are acknowledged with members from IHEP AGN Group. JMW thanks the 
support by the National Key R\&D Program of China through grant-2016YFA0400701 and -2020YFC2201400 by 
NSFC-11991050, -11991054, -11833008, -11690024, and by grant 
No.\,QYZDJ-SSW-SLH007 and No.\,XDB23010400. LCH was supported by the NSFC
(11721303, 11991052, 12233001), the National Key R\&D Program of China (2022YFF0503401), 
and the China Manned Space Project (CMS-CSST-2021-A04, CMS-CSST-2021-A06).
YFY is supported by National SKA Program of China No. 2020SKA0120300 and
National Natural Science Foundation of China (Grant No. 11725312).

\appendix\label{app:cav}
\section{Other processes of cavity formation}
In order to form a cavity, the sMBH-outflows should work to overcome possible barriers.
First, the outflows work against the gravitational energy between the sMBH and the cavity gas.
This means $\Delta E_{1}=GM_{\rm s}\Delta M_{\rm cav}/R_{\rm cav}$, where 
$\Delta M_{\rm cav}=\frac{4\pi}{3}R_{\rm cav}^{3}n_{\rm e}m_{\rm p}$. We have the timescale
of cavity formation
\begin{equation}
\Delta t_{1}=31.1\,\eta_{0.1}^{-1/3}\alpha_{0.1}^{-2/3}r_{1}^{7/6}q_{\bar{5}}^{1/3}
             \left(\frac{M_{6}}{4}\right)\,{\rm s},
\end{equation} 
where $\eta_{0.1}=\eta_{\rm out}/0.1$.
Second, the outflows should overcome the SMBH binding energy, and
$\Delta E_{2}=\Delta M_{\rm cav}(GM_{\rm p}/R_{\bullet}^{2})R_{\rm cav}$. The timescale is given by
\begin{equation}
\Delta t_{2}=31.9\,\eta_{0.1}^{1/3}\alpha_{0.1}^{-4/3}r_{1}^{11/6}
             q_{\bar{5}}^{2/3}\left(\frac{M_{6}}{4}\right)\,{\rm s}.
\end{equation}
The above estimations show that the outflows can easily make a cavity. Third, the outflows 
should work against the gas pressure of the SMBH-ADAF, that is  
$\Delta E_{3}=\frac{4\pi}{3}R_{\rm cav}^{3}P_{\rm gas}$ in the main text 
\citep[e.g.,][]{McNamara2007,Fabian2012}. Comparing the three cases, we find that the third 
case can make a much large cavity than the other two.

\end{document}